\documentstyle[twoside,fleqn,espcrc2,fps]{article}

\newcommand{\AmS}{{\protect\the\textfont2
  A\kern-.1667em\lower.5ex\hbox{M}\kern-.125emS}}

\def\lsi{\raise0.3ex\hbox{$<$\kern-0.75em\raise-1.1ex\hbox{$\sim$}}}
\def\gsi{\raise0.3ex\hbox{$>$\kern-0.75em\raise-1.1ex\hbox{$\sim$}}}

\newcommand{\R}{{\kern+.25em\sf{R}\kern-.78em\sf{I} 
  \kern+.78em\kern-.25em}}
\newcommand{\C}{{\kern+.25em\sf{C}\kern-.50em\sf{I} \kern+.50em\kern-.25em}}
\newcommand{\eps}{\epsilon}
\newcommand{\be}{\begin{equation}}
\newcommand{\ee}{\end{equation}}
\newcommand{\bea}{\begin{eqnarray}}
\newcommand{\eea}{\end{eqnarray}}

\newcommand{\la}{\langle}
\newcommand{\ra}{\rangle}
\newcommand{\bd}{\begin{displaymath}}
\newcommand{\ed}{\end{displaymath}}

\title{Random Matrix Theory and the Spectra of Overlap Fermions
\thanks{Talk presented by S. Shcheredin at Lattice 2003.
\newline \hspace*{0.9mm} Preprint HU-EP-03/55, DESY 03-127, SFB/CPP-03-30}}
\author{S. Shcheredin
\address{ Institut f\"{u}r Physik, Humboldt Universit\"{a}t zu Berlin,
Newtonstr. 15, D-12489 Berlin, Germany \\ 
$^{{\rm b}}$ NIC/DESY Zeuthen, Platanenallee 6, D-15738 Zeuthen, Germany}
, W. Bietenholz $^{{\rm a}}$, T. Chiarappa $^{{\rm b}}$,
K. Jansen $^{{\rm b}}$ and K.-I. Nagai $^{{\rm b}}$
}

\begin{document}

\begin{abstract}

The application of Random Matrix Theory to the Dirac operator
of QCD yields predictions for the
probability distributions of the lowest eigenvalues.
We measured Dirac operator spectra using massless overlap fermions 
in quenched QCD at topological
charge $\nu = 0,$ $\pm 1$ and $\pm 2$, and found agreement with
those predictions --- at least for the first non-zero eigenvalue ---
if the volume exceeds about $(1.2~{\rm fm})^{4}$.

\vspace*{-5mm}

\end{abstract}

\maketitle

In QCD with $N_{f}$ massless quark flavors, chiral symmetry is 
assumed to be broken 
spontaneously as $SU(N_{f})_{R} \otimes SU(N_{f})_{L} \to SU(N_{f})_{R+L}$,
which generates $N_{f}^{2}-1$ Goldstone bosons in the coset space
$SU(N_{f})$. They pick up a mass --- so they can be identified with
light mesons --- in the case of small quark masses.
Their dynamics is described by chiral perturbation theory
as a low energy effective model of QCD.
To the lowest order in the momenta and masses, the effective Lagrangian reads
\be
{\cal L}[U] = \frac{F_{\pi}^{2}}{4} {\rm Tr} \, \left[ \partial_{\mu} U
\partial_{\mu} U \right] - \frac{\Sigma}{2} {\rm Tr} \left[
{\cal M} ( U + 
U^{\dagger} ) \right]
\ee
where $U(x) \in SU(N_{f})$ and 
${\cal M}$ is the (diagonal) quark mass matrix.
The low energy constants $F_{\pi}$ and $\Sigma$
appear here as free parameters. In a box of size $V = L^{4}$ the relations
$\Lambda_{QCD}^{-1} \ll L \ll m_{\pi}^{-1}$ characterize
the {\em $\eps$-regime} \cite{GasLeu}.

In that regime we consider the eigenvalues (EVs) $i\lambda_{n}$ of the 
Dirac operator, resp.\ the dimensionless variable 
$z_{n} = \lambda_{n} V \Sigma$. The spectral density is defined
by $\rho (\lambda ) = \la \sum_{n} \delta (\lambda - \lambda_{n}) \ra$,
and in particular the {\em microscopic spectral density} 
$\rho_{s}$ is given by
\be  \label{micro}
\rho_{s}(z) = \ ^{\lim}_{V \to \infty} \frac{1}{\Sigma V} 
\rho( \frac{z}{\Sigma V}) \ , \quad z = \lambda V \Sigma \ .
\ee
In the sectors of topological charge $\pm \nu$ it can be decomposed as
\be  \label{specdens}
\rho_{s}^{(\nu )}(z) = \sum_{k \geq 1} p_{k}^{(\nu )}(z) \ ,
\ee
where $p_{k}^{(\nu )}(z)$ is the probability distribution of the $k$-th
EV (excluding the zero EV). 
The application of Random Matrix Theory (RMT) to QCD
provides explicit predictions for the
functions $p_{k}^{(\nu )}(z)$ in the $\eps$-regime \cite{lowEV}.

To test these predictions on the lattice, we simulated quenched QCD
with the Wilson gauge action and
the Neuberger overlap operator $D_{\rm ov}$ 
at zero quark mass \cite{BJS}. 
At $\beta =6$ resp.\ $5.85$ the mass of the Wilson kernel, 
$-\mu$, was fixed by $\mu =1.4$ resp.\ $1.6$. 
In contrast to the continuum, where the EVs are
imaginary, the EVs of $D_{\rm ov}$ lie on a circle in $\C$ with center 
and radius $\mu$. For comparison to the continuum predictions we mapped
this circle stereographically onto the imaginary axis.

The functions $p_{k}^{(\nu )}(z)$ depend significantly on $\vert \nu \vert$;
for instance, the peak of $p_{1}^{(\nu )}(z)$ moves to larger values
of $z$ as $\vert \nu \vert$ increases. This effect could not be seen
convincingly, however, in simulations with staggered fermions 
(at least for the couplings investigated so far) \cite{stagger}.
On the other hand, previous studies with Ginsparg-Wilson fermions were 
consistent with the RMT predictions in QED$_{2}$ \cite{Graz},
QED$_{4}$ \cite{QED} and in QCD on a $4^{4}$ lattice \cite{Bern}.
QCD studies on larger lattices were presented in Refs.\ 
\cite{BJS,TSPW}.

\begin{figure}[hbt]
\vspace*{-12mm}
\def\fpsangle{270}
\epsfxsize=42mm
\fpsbox{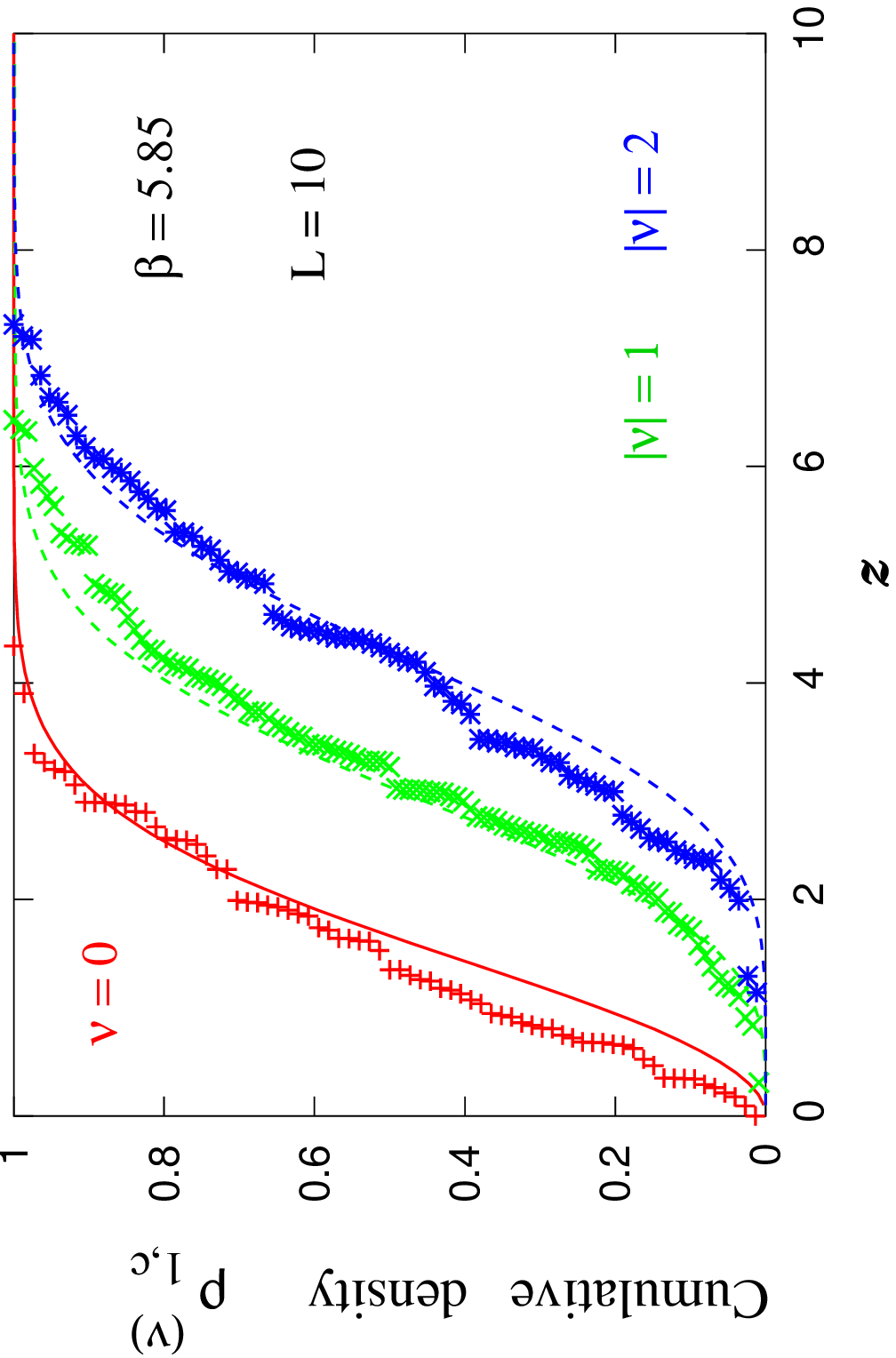}
\def\fpsangle{270}
\epsfxsize=42mm
\fpsbox{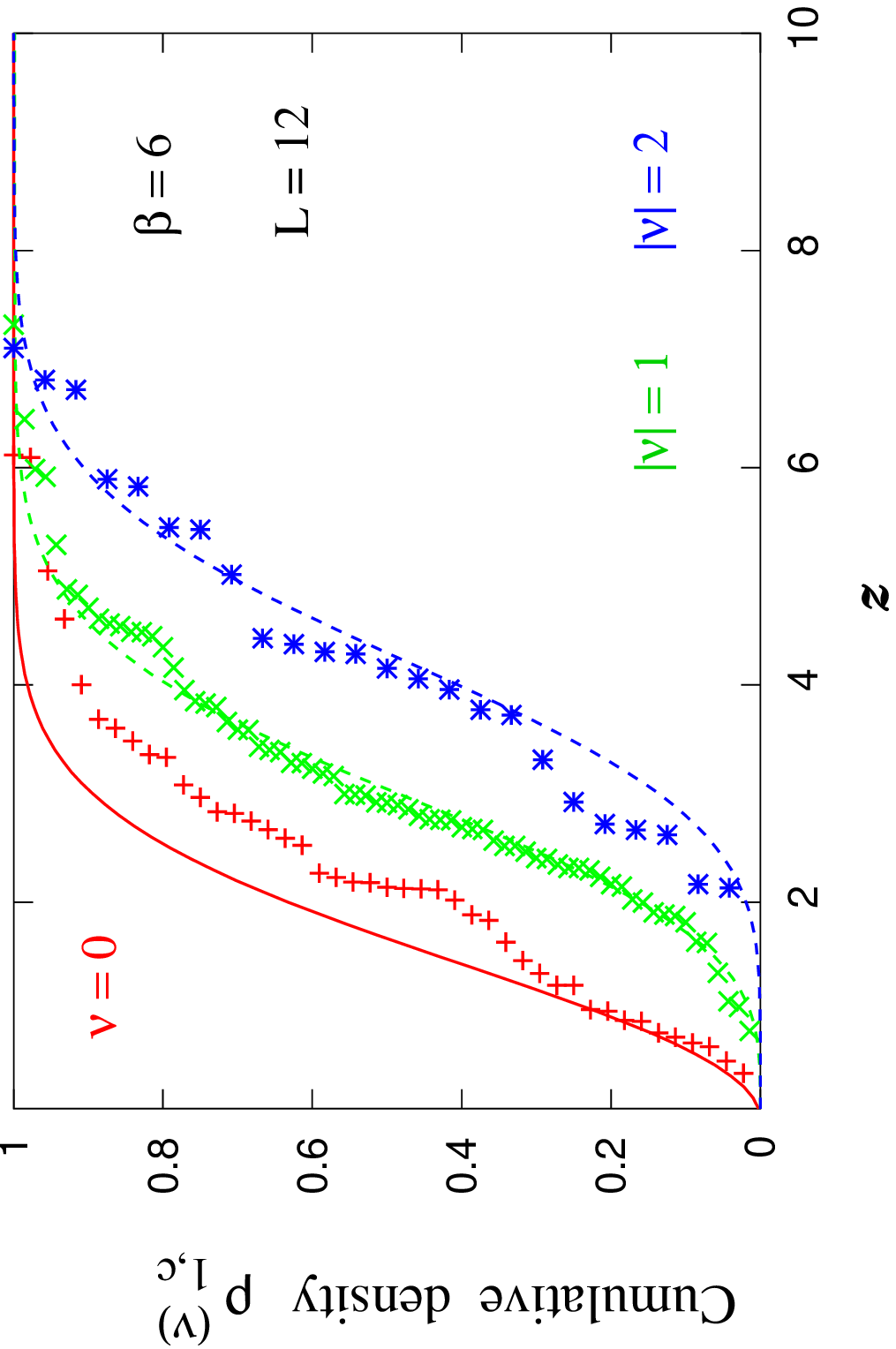}
\def\fpsangle{270}
\epsfxsize=42mm
\fpsbox{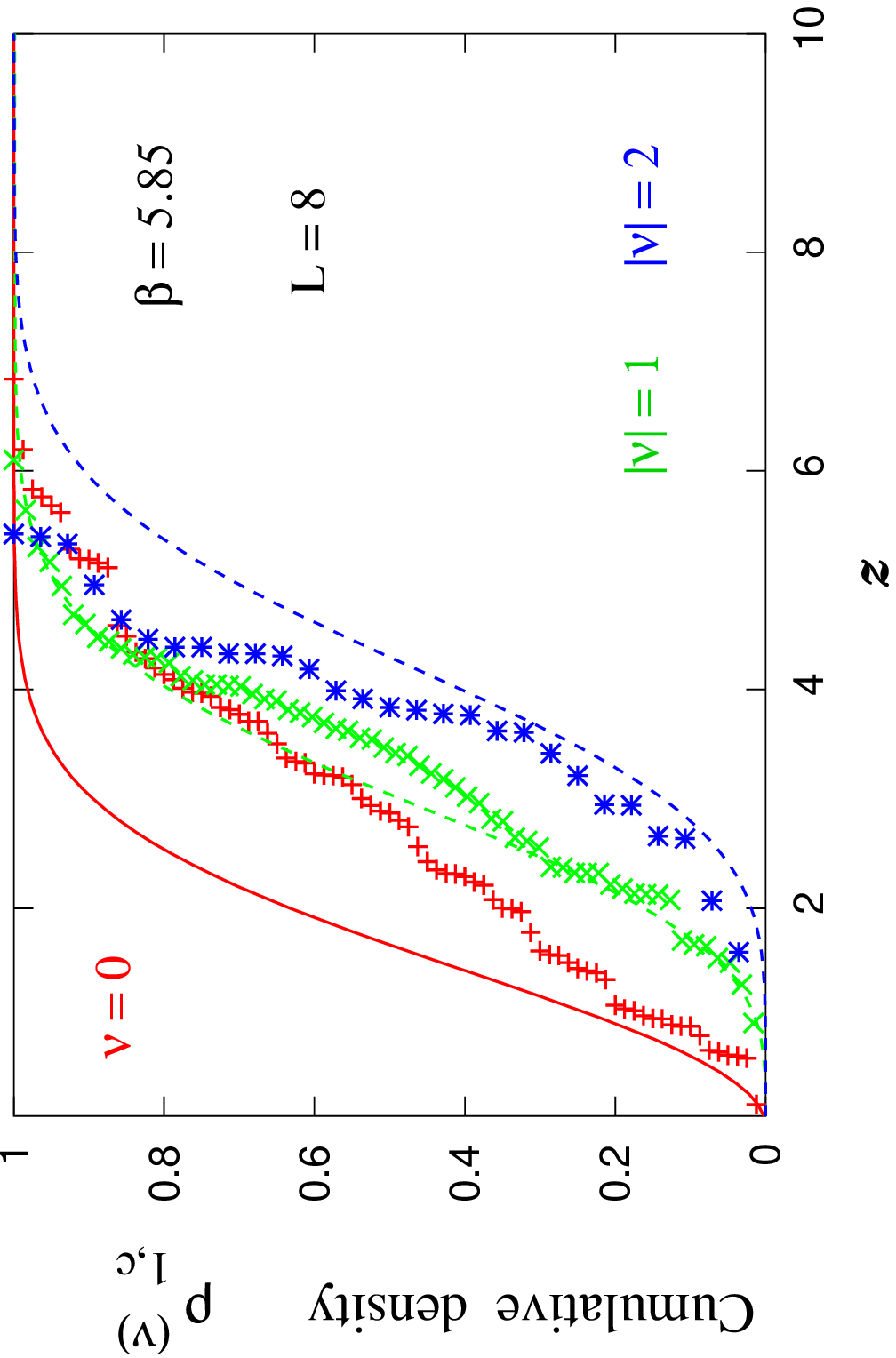}
\vspace*{-9mm}
\caption{\it{The cumulative density of the first non-zero EV in the sectors
$\vert \nu \vert =0,1$ and $2$. We compare the RMT predictions (curves)
to the lattice data on volumes $(1.23~{\rm fm})^{4}$ (top), 
$(1.12~{\rm fm})^{4}$ (center) and $(0.98~{\rm fm})^{4}$ (bottom).
The quality of agreement increases with the volume.}}
\label{cumdens}
\end{figure}

Usually the test of RMT predictions was done by comparing the 
functions $p_{k}^{(\nu )}(z)$ to the histograms of lattice data.
To avoid the arbitrariness of the bin size in that procedure, 
we compared instead the cumulative density
\be
\vspace*{-1mm}
\rho_{k,c}^{(\nu )}(z) = 
\frac{\int_{0}^{z} \rho_{k}^{(\nu )}(z') \, dz'}
{\int_{0}^{\infty} \rho_{k}^{(\nu )}(z') \, dz'} \ .
\ee
In Fig.\ \ref{cumdens} we show the RMT curves for $\rho_{1,c}^{(\nu )}(z)$,
$\vert \nu \vert = 0,$ 1 and 2, compared
to data from the following lattices: $10^{4}$ at $\beta =5.85$,
$12^{4}$ at $\beta =6$ and $8^{4}$ at $\beta =5.85$. The corresponding
physical volumes are $\approx (1.23~{\rm fm})^{4}$, $(1.12~{\rm fm})^{4}$
and $(0.98~{\rm fm})^{4}$. We see that the $k=1$ EV tends to agree
quite well with RMT, {\em if} the volume exceeds about $(1.2~{\rm fm})^{4}$.
It disagrees, however, if the physical volume is too
small; then the data do not distinguish any more the different
topological sectors, which is consistent with Ref.\ \cite{Bern}.

This comparison involves
$\Sigma$ as the only free parameter, which was
optimized for agreement with RMT. In
$(1.23~{\rm fm})^{4}$ and $(1.12~{\rm fm})^{4}$ we obtained very similar
values, $\Sigma = (253~{\rm MeV})^{3}$ resp.\
$(256~{\rm MeV})^{3}$. These values are consistent with the literature,
although $\Sigma$ is expected to diverge logarithmically at large $V$
in quenched QCD \cite{loga}. Further volumes and higher statistics would
be required to verify this behavior; our statistics was too small
to associate error bars with the above values for $\Sigma$.

\begin{figure}[hbt]
\vspace*{-7.5mm}
\def\fpsangle{270}
\epsfxsize=43mm
\fpsbox{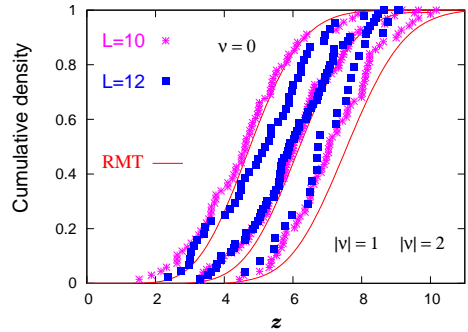}
\vspace*{-9mm}
\caption{\it{The cumulative density of the second non-zero EV 
at 
$\vert \nu \vert =0,\, 1,\, 2$. We compare the RMT predictions (curves)
to the lattice data in the volumes $(1.23~{\rm fm})^{4}$ (stars) and
$(1.12~{\rm fm})^{4}$ (squares). The agreement is worse than
in Fig.\ \ref{cumdens}, since the relevant values
of $z$ are larger.}}
\label{secondEV}
\vspace*{-8mm}
\end{figure}

The range of $z$ where the predictions are
compatible with the RMT curves raises gradually as the volume is enlarged.
For $V < (1~{\rm fm})^{4}$ this range is so short that it doesn't even
capture the peak of the first non-zero EV, hence the RMT prediction
is not applicable in that regime. However, our largest volume is still not 
sufficient to capture well the $k=2$ EV; we show results for our larger
two volumes in Fig. \ref{secondEV}.

\begin{figure}[hbt]
\def\fpsangle{270}
\epsfxsize=43mm
\fpsbox{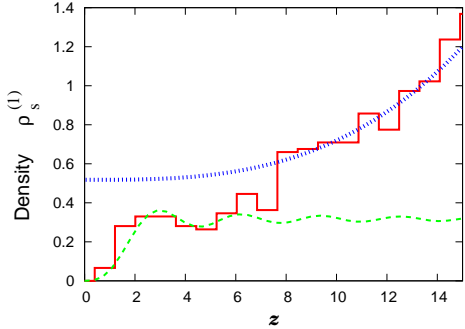}
\vspace*{-9mm}
\caption{\it{The spectral density of the low lying EV on a lattice of
volume $(1.12~{\rm fm})^{4}$. The histogram of the lattice data roughly
follows the RMT prediction (dashed line)
up to the second peak, before turning into the cubic bulk behavior
(dotted line).}}
\label{rho_s}
\vspace*{-8mm}
\end{figure}

Next we present results for the full spectral density
$\rho_{s}^{(\nu )}(z)$ of eq.\ (\ref{micro}) (summed over $k$). 
Fig.\ \ref{rho_s} shows a histogram at $\vert \nu \vert =1$
on the $12^{4}$ lattice, which roughly
follows the RMT prediction up to the second peak, before turning into the
bulk behavior $\rho_{s}^{(\nu )}(z) = c_{0} + c_{1} z^{3}$.

\begin{figure}[hbt]
\vspace*{-11mm}
\def\fpsangle{270}
\epsfxsize=46mm
\fpsbox{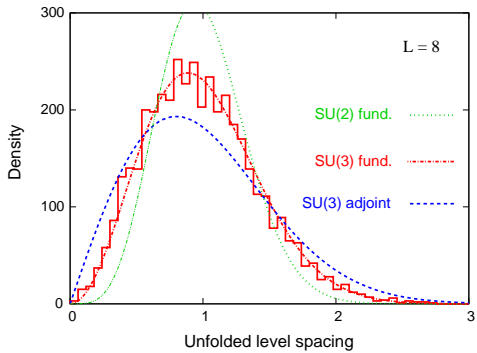}
\vspace*{-9mm}
\caption{\it{The unfolded level spacing distribution
on a $8^4$ lattice at $\beta =5.85$. The histogram follows
closely the curve predicted for the gauge group $SU(3)$
in the fundamental representation.}}
\label{unfold}
\vspace*{-9mm}
\end{figure}

We also studied the ``unfolded level spacing distribution'' which was
considered earlier on a $4^{4}$ lattice \cite{SCRI}.
There we found good agreement with the behavior specific for $SU(3)$
in the fundamental representation that we were using. In this case,
the results agree well with the prediction on all the three lattice sizes
that we considered; the lower limit for the volume does not
seem to affect this quantity, see Fig.\ \ref{unfold}. 
However, this type of ``unfolding'' only 
keeps track of the relative order of the EV in the ensemble, hence it
is not sensitive to the energy scale and not directly physical.\\

In view of the future numerical exploration of the $\eps$-regime
by means of Ginsparg-Wilson fermions, our main conclusions are
\begin{itemize}

\vspace*{-2mm}
\item The volume should be larger than about $(1.2 ~ {\rm fm})^{4}$,
which might just reflect that chiral perturbation theory
is only applicable in the confinement phase.
This observation can also be viewed as an empirical determination of the
so-called Thouless energy.

\vspace*{-2mm}
\item Once we are in the right regime, the RMT predictions for the low lying
EVs are confirmed in some range $ z= 0 \dots z_{\rm crit}$,
where $z_{\rm crit}$ grows with the volume.

\vspace*{-2mm}
\item The latter means that particularly in the topologically
neutral sector there is a non-negligible density of very small non-zero 
EVs. Such EVs are dangerous for the measurement of physical observables;
they lead to the requirement of a huge statistics \cite{nuzero}. 

\end{itemize}
\vspace*{-1mm}

In the $\eps$-regime, observables tend to be strongly $\vert \nu \vert$ 
dependent 
\cite{LeuSmi}, hence they should be measured at fixed $\vert \nu \vert$.
The last observation now implies that such measurements
should beware of $\nu =0$.

\end{document}